\documentclass[12pt,journal,draftcls,letterpaper,onecolumn]{IEEEtran}

\usepackage{graphicx}  
\usepackage{amsmath}
\usepackage{amssymb}

\usepackage{algorithm}
\usepackage{algorithmic}
\usepackage{xcolor}
\usepackage{color}
\usepackage{makecell}
\newtheorem{theorem}{Theorem}

\newtheorem{definition}[theorem]{Definition}

\newtheorem{proposition}[theorem]{Proposition}

\newtheorem{assumption}[theorem]{Assumption}

\definecolor{darkgreen}{rgb}{0.1,0.5,.5}
\definecolor{darkred}{rgb}{0.8,0,0}
\definecolor{teal}{rgb}{0.05,0.32,0.41}
\definecolor{darkblue}{rgb}{0,0.0,0.5}
\definecolor{blackgreen}{rgb}{0,0.4,0}
\definecolor{purple}{rgb}{0.5,0,0.3}
\definecolor{grey}{rgb}{0.7,0.5,0.5}
\definecolor{orange}{rgb}{0.6,0.4,0.1}

\begin{document}
%
\title{A New Signature Scheme Based on Punctured Reed--Muller Code With Random Insertion}
%
%
\author{Wijik Lee, Young-Sik Kim, and Jong-Seon No
\thanks{Y.-S. Kim is the corresponding author.}
\thanks{W. Lee and J.-S. No are with the Department of Electrical and Computer
Engineering, INMC, Seoul National University, Seoul 08826, Korea, e-mail:
leewj422@ccl.snu.ac.kr, jsno@snu.ac.kr.}
\thanks{Y.-S. Kim is with the Department of Information and Communication Engineering, Chosun University, Gwangju 61452, Korea, email: iamyskim@chosun.ac.kr.}

}

%
%
%
\markboth{Submission}{Regular Paper}



\maketitle

\begin{abstract}
In this paper, we propose a new signature scheme based on a punctured Reed--Muller (RM) code with random insertion, which improves the Goppa code-based signature scheme developed by Courtois, Finiasz, and Sendrier (CFS). The CFS signature scheme has certain drawbacks in terms of scaling of the parameters and a lack of existential unforgeability under adaptive chosen message attacks (EUF-CMA) security proof. Further, the proposed modified RM code-based signature scheme can use complete decoding, which can be implemented using a recursive decoding method, and thus syndromes for errors larger than the error correctability can be decoded for signing, which improves the probability of successful signing and reduces the signing time. Using the puncturing and insertion methods, the proposed RM code-based signature scheme can avoid some known attacks for RM code-based cryptosystems. The parameters of the proposed signature scheme such as error weight parameter $w$ and the maximum signing trial $N$, can be adjusted in terms of signing time and security level, and it is also proved that the proposed signature scheme achieves EUF-CMA security.
\end{abstract}


\vspace{10pt}
\section{Introduction}
As quantum computers continue to be developed, conventional public key cryptosystems such as an RSA cryptosystem and an elliptic curve cryptosystem are expected to be broken in the near future through the application of efficient quantum algorithms. 
Thus, there have been several studies on the development of robust cryptosystems that are immune to attacks by quantum computers, namely, post-quantum cryptography. Code-based cryptography is considered as one of the possible candidates for post-quantum cryptography. In 1978, McEliece \cite{mcelice78} first proposed a code-based cryptosystem using a generator matrix of binary Goppa code, and later Niederreiter \cite{Niederreiter} suggested another version of code-based cryptosystem using a parity check matrix based on a syndrome decoding problem known as an NP-complete problem \cite{NPHard}. Many code-based cryptosystems have since been proposed by replacing the binary Goppa code with other error correcting codes. However, no valid code-based digital signature scheme were proposed for more than two decades after that point. 

In 2001, Courtois, Finiasz, and Sendrier introduced the first code-based digital signature scheme, called the Courtois, Finiasz, and Sendrier (CFS) signature scheme \cite{CFS}. The CFS signature scheme is based on Niederreiter cryptosystem. In the signing process of the CFS scheme, the hash of message $M$, $h(h(M)|i)$, is generated, and is considered as a syndrome of the given code, where $h({\cdot})$ is a cryptographic hash function from $\{0,1\}^*$ to $\{0,1\}^{n-k}$, and $i$ is a counter value used to adjust the hashed message corresponding to the syndrome for a valid error. Therefore, to generate a valid signature, we need to search the corresponding error vector $e$ whose Hamming weight is less than or equal to the error correctability $t$ of the code such that $He^T = h(h(M)|i)$. This error vector and counter $i$ are the signature of the given message $M$. Clearly, finding $e$ is a syndrome decoding problem, and it is known that, to find a valid signature $e$, $t!$ trials (increase $i$, recalculate $h(h(M)|i)$, and apply syndrome decoding to find $e$) are expected on average for the case of Goppa codes in the CFS signature scheme \cite{CFS}, which requires a tremendously large number of signing trials for a large $t$. Therefore, the CFS signature scheme is only applicable for a relatively small $t$, and is thus based on a high rate Goppa code.
However, it is known that the generator matrix of a high rate Goppa code can be distinguished from a random matrix \cite{Distinguish}, and thus
the CFS scheme is not robust or existentially unforgeable against a chosen message attack (EUF-CMA) \cite{Distinguish}.

Thus, there have been many efforts to relieve the security problem of the CFS signature scheme. One approach is to adopt other codes in place of the Goppa code. For example, low-density generator matrix (LDGM) \cite{LDGM} and convolutional code-based signature schemes have been proposed \cite{Escher}. However, the LDGM code-based signature scheme was recently proved to be insecure \cite{LDGMattack}. After collecting a large number of signatures, attackers can find the correlation of signatures and can then decompose the public key $H'=SHQ$ into the private keys $S$, $H$, and $Q$. Another approach to relieve the security problem of the CFS signature scheme is utilizing a modified Goppa code. Because a high rate Goppa code-based signature scheme is not secure against EUF-CMA \cite{Distinguish}, they proposed a modified CFS scheme using an $(n,k-1)$ expurgated Goppa code to evade the Goppa code distinguishing problem \cite{Punctured}. In general, the CFS signature scheme has a small value of $t$, which causes a vulnerability to birthday attacks \cite{Birthday}. It is noted that if we can use complete decoding, the decoding for errors larger than $t$ can be possible, which improves the security and successful signing probability.

In this paper, we propose a new variant of the CFS signature scheme based on punctured Reed--Muller (RM) code with random insertion. The modified RM code can perform complete decoding by utilizing a well-known and efficient recursive decoding \cite{Closest}, \cite{Recursive}, called closest coset decoding, that is, for a given received vector, the closest codeword can be found. The closest coset decoding method does not guarantee an exact error correction, but finds an error vector (coset leader in the standard array) corresponding to the syndrome. However, the exact error correction is not essential for signing in code-based signature schemes, but we need to find the error vector with the smallest Hamming weight in the coset corresponding to the syndrome. In this respect, the RM code-based signature scheme can be considered as a solution to the small $t$ constrained problem of the Goppa code-based signature scheme. Further, the proposed signature scheme can compromise the signing time and security level by adjusting the allowable maximum Hamming weight of error vectors, called the error weight parameter $w = t + \delta$.

However, the simple replacement of Goppa code with RM code in the CFS signature scheme results in vulnerability to several attacks. The RM code-based McEliece cryptosystem \cite{RM} is insecure owing to Minder--Shokrollahi attack \cite{RMattack} and Chizhov--Borodin attack \cite{RMattack2}. With these two attacks, the private keys $S$, $G$, and $Q$ can be revealed from the public key $G'=SGQ$. These attacks can be similarly applied to the RM code-based signature scheme. It is shown herein that punctured RM codes with random insertion can be secure against these attacks, and an optimal puncturing scheme for preventing Minder--Shokrollahi and Chizhov--Bordin attacks is proposed \cite{modifiedRM}. In addition, it is also shown that the punctured RM code with random insertion is secure from a square code attack \cite{RMsquare}, which can distinguish randomly inserted columns from the modified generator matrix. In this paper, it is also proved that the proposed modified RM code-based signature scheme is EUF-CMA secure under the assumption that the parity check matrix of the modified RM code is not distinguishable from a random matrix.

The remainder of this paper is organized as follows. In Section II, we introduce conventional CFS signature schemes, the RM code, and the punctured RM code with random insertion. In Section III, we propose a modified CFS signature scheme based on the punctured RM code with random insertion. Then, the security of the proposed signature scheme is presented in Section IV. Finally, Section V provides some concluding remarks regarding this research.

\vspace{10pt}
\section{Conventional Code-Based Signature Scheme} \label{sec_preliminaries}

In this section, we introduce the conventional code-based signature scheme (i.e., CFS signature scheme \cite{CFS}), the RM codes, and its puncturing method with random insertion.

\subsection{CFS Signature Scheme}

Courtois, Finiasz, and Sendrier proposed the first practical code-based signature scheme, called CFS signature scheme. It is based on the Niederreiter cryptosystem and consists of three stages, namely, key generation, signing, and verification, as follows.

\vspace{0.1in}

{\bf Key Generation:} Choose a parity check matrix $H$ of an $(n,k)$ binary $t$-error correcting Goppa code with a decoding algorithm $\gamma$. The decoding algorithm $\gamma$ will produce an error vector $e$ with Hamming weight of less than or equal to the error correctability $t$ if it exists, or output $\perp$, otherwise.
Let $S$ be an $(n-k)\times (n-k)$ scrambling matrix, and $Q$ be an $n\times n$ permutation matrix. Construct the public key $H'=S H Q$, where $S$, $Q$, and $H$ are the private keys.

\vspace{0.1in}

{\bf Signing:} To sign a message $M$,
\begin{list}{ }{\setlength{\leftmargin=15mm \labelwidth=1mm}}
\item 1) $i\leftarrow i+1$
\item 2) $e' = \gamma (S^{-1}h(h(M)|i))$
\item 3) if $e'$ is $\perp$, go to Step 1).
\item Output $(M,e = e'(Q^{-1})^T, i)$
\end{list}

\vspace{0.1in}

{\bf Verification:}
\begin{list}{ }{\setlength{\leftmargin=15mm \labelwidth=1mm}}
\item Compute $\hat{s} = H'e^T$ and $s=h(h(M)|i)$.
\item The signature is valid if $s$ and $\hat{s}$ are equal.
\end{list}
\vspace{0.1in}

The security of the CFS signature scheme is based on the hardness of solving a syndrome decoding problem, which is known as an NP-complete problem \cite{NPHard}.

\begin{definition}[Syndrome decoding problem]
{Given an $r\times n$ parity check matrix $H$, a syndrome $s\in\{0,1\}^r$, and an error correctability $t > 0$, find an error vector $e$ in $He^T=s$ with Hamming weight of less than or equal to $t$.
}
\end{definition}

In the signing process of the CFS signature scheme, the hashed message $h(h(M)|i)$ with counter $i$ is treated as a syndrome. However, it is known that the ratio of successfully decodable syndromes is only $1/t!$ for the case of the CFS signature scheme with binary $t$-error correcting Goppa code \cite{CFS}. Therefore, to obtain a valid signature, we need to search a valid error vector by carrying out $t!$ trial decodings on average, and thus $t$ should be small.

In the proposed signature scheme, we consider another decoding method, called a complete decoding problem, which is finding a nearest codeword to the received vector in the vector space.

\begin{definition}[Complete decoding problem \cite{CFS}]
{Given an $r\times n$ parity check matrix $H$ and a syndrome $s\in\{0,1\}^r$, find an error vector $e$ with the minimum Hamming weight in $\{e|He^T=s\}$.
}
\end{definition}

Complete decoding problem is known as the most difficult computational problem in decoding \cite{code-based}. Complete decoding makes it possible to find an error vector with Hamming weight of greater than $t$ for the given syndrome at the cost of large computational complexity \cite{Closest}, \cite{Recursive}. However, when we apply the complete decoding to the signing in the CFS signature scheme with binary Goppa code, there is a limitation in that the value of $\delta$ in $w=t+\delta$ cannot be sufficiently large.

In addition, there are some security drawbacks to the CFS signature scheme: (i) the parity check matrix of high rate Goppa code can be distinguished from a random matrix, and thus the CFS signature scheme is insecure under the EUF-CMA, and (ii) it has poor scaling of the parameters based on the security as in the following description. The error correcting parameter $t$ needs to be small because the number of operations required for the generation of valid signature is significantly dependent on $t$, that is, $t! t^2m^3$, where $n-k=tm$. The public key size of the CFS scheme is $(n-k)n = tm2^m$, and it is known that decoding attacks require $A = 2^{tm/2}$ operations. Thus the decoding attack complexity $A$ is only a polynomial function of the key size with small power, that is, $A \approx {\rm key size}^{t/2}$. Therefore, because $t$ should be kept as a relatively small value of up to $12$ to reduce successful signing time, we need to significantly increase the key size itself for higher security.

\subsection{RM Code and Its Modification}
In this paper, we proposed the CFS signature scheme using the modified RM codes. Because the RM code and its modified one can be decoded through complete decoding, they can improve the security drawback of the CFS signature scheme with binary Goppa code by extending the error correctability $t$ to the error weight parameter $w$.

\subsubsection{RM Code}
The Reed--Muller code, RM$(r,m)$, of order $r$ is a linear code defined by Boolean functions for any integers $m$ and $r$ with $0 \le r \le m$. A Boolean function of $m$ variables can be evaluated on $2^m$ different positions, and thus the RM code has codewords of length $2^m$. The RM code, RM$(r,m)$, is a set of codewords obtained by evaluating all Boolean functions of a degree of less than or equal to $r$. Thus, the length $n$, dimension $k$, and minimum distance $d$ of RM$(r,m)$ are given as
\begin{equation*}
n=2^m, k=\sum_{i=0}^{r}{m\choose i}, d=2^{m-r}.
\end{equation*}

\subsubsection{Complete Decoding of RM Code}
Because the recursive decoding of the RM code can find the coset leader of the received vector, this decoding can be considered as closest coset decoding \cite{Recursive}, \cite{Closest}. In fact, closest coset decoding is the same as complete decoding. Therefore, the RM code-based CFS signature scheme is worth considering. However, simply modifying the CFS signature scheme by replacing it with an RM code is easily broken by the well-known Minder-Shokrollahi and Chizhov-Borodin attacks, which are used in the RM code-based McEliece cryptosystem \cite{modifiedRM}. With these attacks, the private keys $S$, $H$, and $Q$ of the cryptosystem can be obtained from the public key $H'=SHQ$. Therefore, we need to modify the RM code structure to achieve security under known attacks, while maintaining the complete decodable property of the RM code \cite{modifiedRM}.

\subsubsection{Puncturing RM Code with Random Insertion}
In \cite{modifiedRM}, the punctured RM code with random insertion is introduced to construct a secure RM code-based public key cryptosystem. Similarly, the puncturing and random insertion methods can be applied to an RM code-based signature scheme. In fact, the puncturing of a generator matrix $G$ is equivalent to row and column deletions of a parity check matrix $H$ \cite{Shortening}. Because a signature scheme uses a parity check matrix $H$, the modified $H$ with row and column deletions and random row insertion will be used for the proposed signature scheme. The modification method of the $(n,k)$ RM code is given as follows.

{\hspace{10pt} {\em {a)} Row and Column Deletions of Parity Check Matrix}}

The systematic forms of generator and parity check matrices are given as
\begin{equation}
G = [I_k|P], H = [P^T|I_{n-k}]
\end{equation}
where $P$ is $k \times (n-k)$ matrix given as
\begin{equation}
P = [p_1 p_2{\cdots}p_{n-k}]
\end{equation}
with column vectors $p_i$ of size $k$.
The generator matrix $G$ can be punctured by deleting columns of matrix $P$. Let $P'$ be a $p$ column deleted matrix from $P$. Then, the generator matrix and parity check matrix of the punctured RM code are given as
\begin{equation}
G_{p} = [I_k|P'], H_{p} = [P'^T|I_{n-k-p}],
\end{equation}
respectively. It can then be easily checked that
\begin{equation}
G_{p} H^T_{p} = 0_{k\times (n-k-p)}.
\end{equation}

{\hspace{10pt} {\em {b)} Modification of Parity Check Matrix with Random Row Insertion}}

The punctured generator matrix $G_{p}$ can be modified by inserting random columns into $P'$. Then, the punctured generator matrix with random column insertion is denoted as $G_{m} = [I_k|P'']$, where $p$ random columns are inserted in $P''$. Then, the generator matrix and its parity check matrix are given as
\begin{equation}
G_{m} = [I_k|P''], H_{m} = [P''^T|I_{n-k}].
\end{equation}
Clearly, we have $G_{m} H^T_{m} = 0_{k\times (n-k)}$.

\vspace{10pt}
\section{New Signature Scheme Using Punctured RM Code with Random Insertion}

In this section, we propose a new code-based signature scheme, which is a modified version of the CFS signature scheme based on punctured RM codes with random insertion. The proposed signature scheme can improve the probability of successful signing, and guarantee EUF-CMA security, which is composed of three stages, namely, key generation, signing, and verification, as follows.

\subsection{Proposed Signature Scheme}
\vspace{0.1in}
{\bf 1) Key Generation}

{\bf 1-1) Puncturing with random insertion:} Let $G$ be a $k\times n$ generator matrix of the RM code, RM$(r,m)$. In this paper, we assume the systematic form of the RM code and $L_D$ is a set of indices of puncturing positions in the parity part $P$ of the systematic form of the generator matrix, which was described for the nonsystematic RM code in Algorithm 1 \cite{modifiedRM}. The procedure for puncturing generator matrix and determining the set $L_D$ is described in Algorithm 1, and two important notations for Algorithm 1 are defined as follows.

\begin{definition}[\cite{modifiedRM}] \label{def_support_x}
The support of a codeword $c\in\;$RM$(r,m)$ is defined as the set of indices $i$ such that $c_i\neq 0$, denoted as supp$(c)$.
\end{definition}

\begin{definition}[\cite{modifiedRM}] \label{def_projection}
Let $c$ be a codeword of $C$ and {$L$} be an index set. Then, {${\rm proj}_L(c)$} is a sub-codeword composed of the components with indices in {$L$} from $c$. In addition, for a linear code $C$, we define {${\rm proj}_L(C) = \{{\rm proj}_L(c)|c\in C\}$}.
\end{definition}

Because the puncturing procedure of the generator matrix of RM code is given in Algorithm 1, the parity check matrix $H$ corresponding to $G$ can be modified.
\begin{algorithm}[!t]
\caption{\cite{modifiedRM} Puncturing procedure of generator matrix}
Input: $k\times n$ generator matrix $G$ of RM code\\
Output: index set $L_D$
\begin{enumerate}
\item[1.] Randomly pick a minimum Hamming weight codeword $x$ from $C$.
\item[2.] Randomly pick a minimum weight codeword $y$ from ${\rm proj}_{{\rm supp}(x)} (C)$.
\item[3.] Choose $p$ such that ${\rm wt}(y) \le p \le 2{\rm wt}(y)$.
\item[4.] Randomly choose the set $L_D$ of indices  such that ${\rm supp}(y) \subseteq L_D$ and $|L_D|=p$.
\end{enumerate}
\end{algorithm}
Some of elements in $L_D$ of Algorithm 1 may be in the information part $I$ of the generator matrix $G=[I|P]$, but we modify the generator matrix into the systematic form such that all elements of $L_D$ should be in the parity part $P$.
Using $L_D$ in Algorithm 1, a modification algorithm of the parity check matrix corresponding to the punctured generator matrix is proposed in Algorithm 2, where the systematic form of the generator matrix $G = [I|P]$ and the parity check matrix $H=[P^T|I]$ are used.

Further, the generator and parity check matrices are row-scrambled and column-permuted to generate the public key in the signature scheme. Thus, without a loss of generality, we can assume that the last $p$ columns of $P$ in $G$ are punctured, and thus we have $L_D = \{n-k-p+1,n-k-p+2,{\cdots},n-k\}$. Therefore, the modification of the parity check matrix is described in Algorithm 2.

\begin{algorithm}[!t]
\caption{Modification of parity check matrix of punctured RM code in systematic form}
Input: $k\times n$ generator matrix $G = [I|P]$ of the systematic form of RM code.\\
Output: modified parity check matrix $H_{m}$.
\begin{enumerate}
\item[1.] Let $L_D = \{n-k-p+1,n-k-p+2,{\cdots},n-k\}$ be an row index set in the systematic form of parity check matrix $H=[P^T|I]$ using $p$ in Algorithm 1.
\item[2.] Replace the last $p$ rows of the parity check matrix $H$ by the binary random vectors $r_i$, $n-k-p+1\le i\le n-k$ denoted as $H_m$, where $$r_{ij}=
\begin{cases}
1, \;\;{\rm for}\;\; j = i+k \\
0, \;\;{\rm for}\;\; n-p+1 \le j \le n, j\neq i+k \\
{\rm random\;selection\;of\;binary\;bits, otherwise}.
\end{cases}$$
\end{enumerate}
\end{algorithm}

\begin{figure}
        \begin{center}
        \includegraphics[width=0.7\columnwidth,keepaspectratio]
        {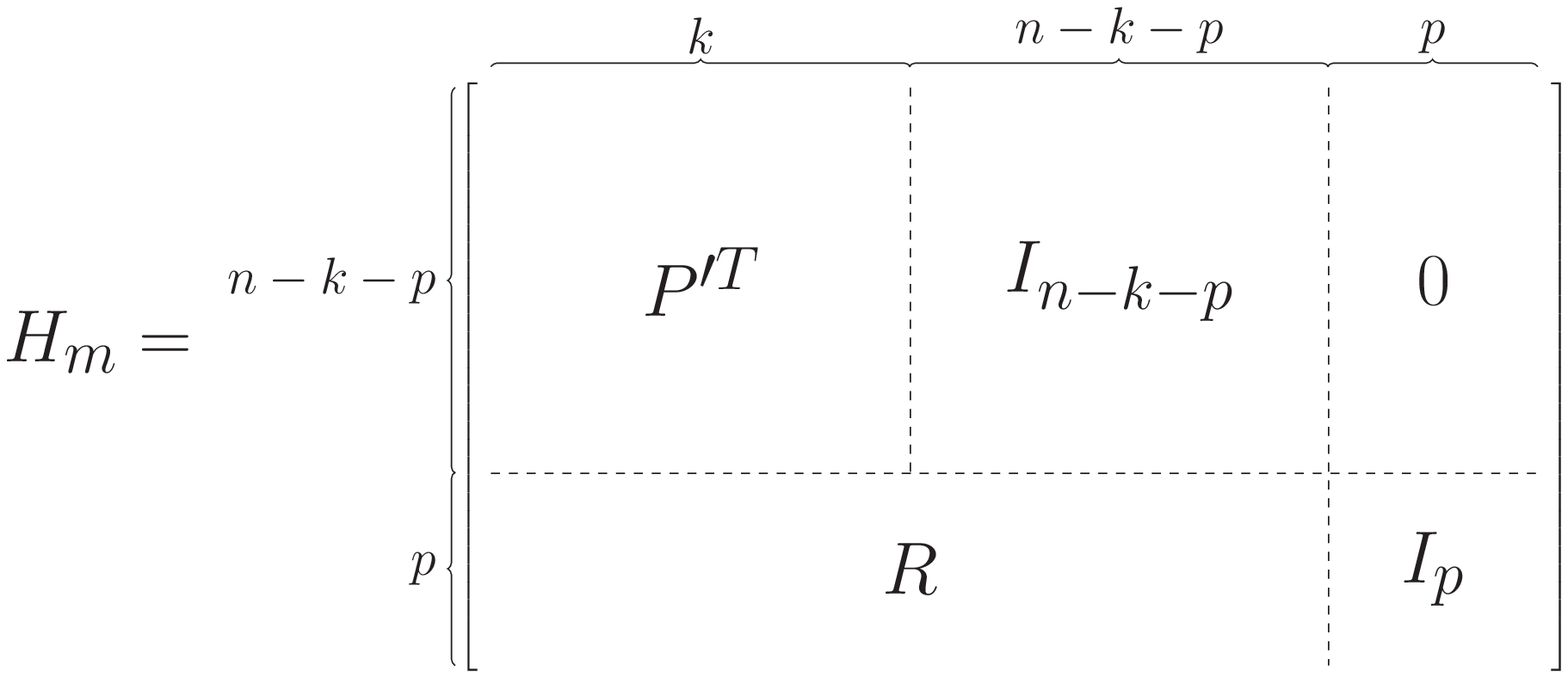}%
        \caption{Modified parity check matrix of the proposed signature scheme.}
        \label{fig_matrix}
        \end{center}
\end{figure}

Then, the modified parity check matrix $H_{m}$ can be described as in Fig.~\ref{fig_matrix},
where $R$ is a $p\times(n-p)$ binary random sub-matrix with row vectors $r_i=(r_{ij})$, $n-k-p+1\le i \le n-k, 1\le j\le n-p$, and $P'$ is the last $p$ column deleted version of $P$.

The deleted and inserted rows are not necessarily the same number as well as the same position but here, we assume that they are the same.

{\bf 1-2) Generation of $S, Q,$ and $H_m$:} Let $S$ be an $(n-k)\times (n-k)$ scrambling matrix and $Q$ be an $n\times n$ permutation matrix. The public key is generated by calculating $H' = SH_{m}Q$, and the private keys are $S, H_{m}$, and Q.

{\bf 2) Signing}

For a message $M$, counter $i$, and hash function $h$, define the syndrome as $s=h(h(M)|i)$, which is the same as that of the CFS signature scheme.

{\bf 2-1) Find the closest coset:} Find the error vector $e$ such that $SHQe^T = s$. Let $e'^T = Qe^T$ and $s' = S^{-1}s$. Then, $He'^T = s'$. Decode the error vector $e'$ by the closest coset decoding.

\begin{figure}
        \begin{center}
        \includegraphics[width=0.9\columnwidth,keepaspectratio]
        {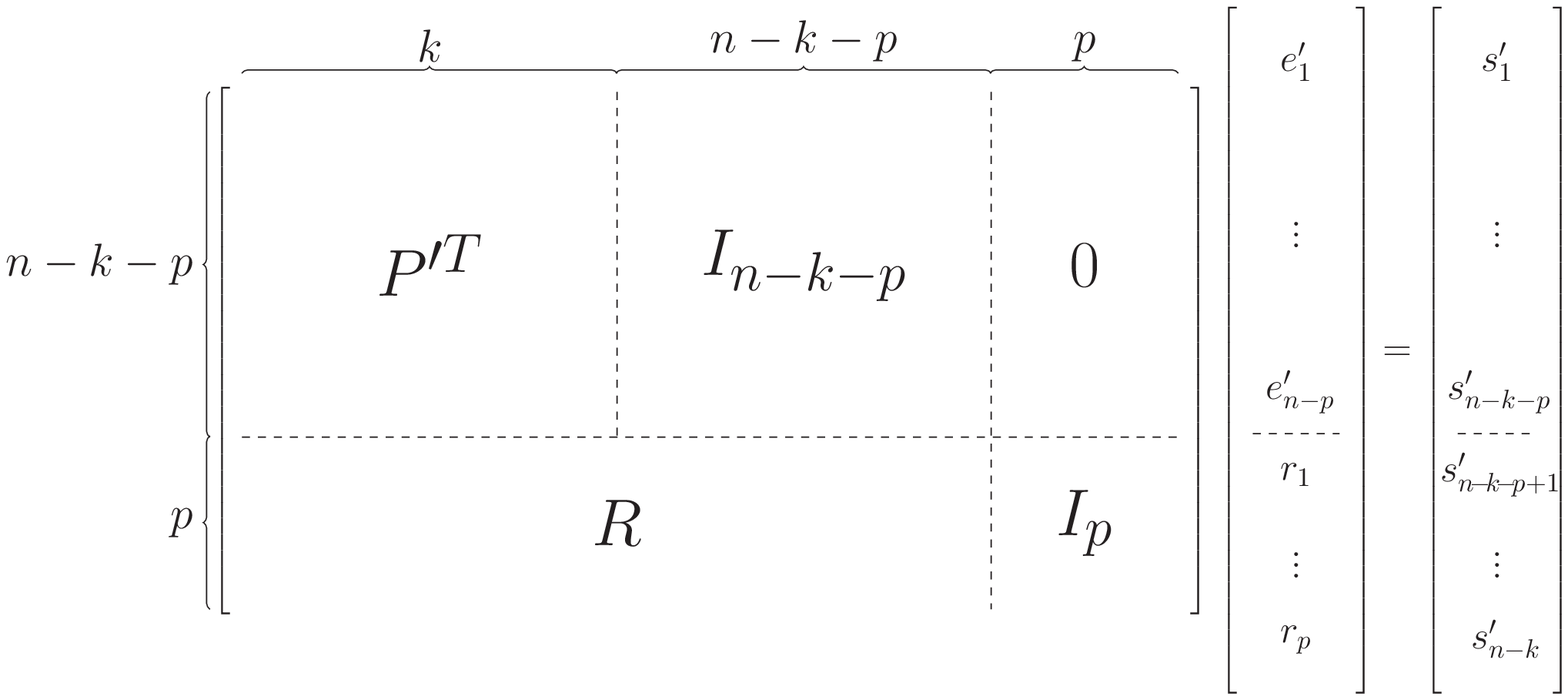}%
        \caption{Signing process of the proposed signature scheme.}
        \label{fig_3}
        \end{center}
\end{figure}

{\bf 2-2) Find the punctured part of the error vector:}
Because the parity check matrix $H$ is random row-deleted and inserted as $H_{m}$, we have to replace the last $p$ elements of $e'$ by $e'_p=[r_1,r_2,{\cdots},r_p]$, denoted as $e' = [e'_{n-p}|e'_{p}]$, such that $H_{m}e'^T=s'$. Let $s'=[s'^T_{n-k-p}|s'^T_{p}]^T,$ where $s'_{n-k-p}$ and $s'_{p}$ denote the first $n-k-p$ and last $p$ elements of $s'$, respectively. Then, $H_{m}e'^T = s'$ can be rewritten as $$\left[\begin{array}{c}[P'^T|I_{n-k-p}] e'^T_{n-p}\\Re'^T_{n-p}+e'^T_p\end{array}\right]=\left[\begin{array}{c}s'_{n-k-p}\\s'_p\end{array}\right].$$ Thus, we have $$e'^T_p = s'_p+Re'^T_{n-p}$$
and thus $e' = [e'_{n-p} | s'_p+Re'_{n-p}]$.

If the Hamming weight of $e'$ is larger than the error weight parameter $w$, then we increase the counter $i$ and apply the signing process again, where $w$ is larger than the error correctability $t$. The maximum number of iterations of the counter $i$ is given as $N$, which will be discussed in the next subsection.

If ${\rm wt}(e') \le w$, compute $e^T = Q^{-1} e'^T$, and the signature $\sigma$ is then given as $\sigma = (M, e, i)$.

{\bf 3) Verification}

Check ${\rm wt}(e) \le w$ and $H'e^{T} = h(h(M)|i)$. If TRUE, then return ACCEPT; else, return REJECT.

\subsection{Preprocessing for Error Weight Parameter}
In the proposed signature scheme, choosing the error weight parameter $w$ is significant for balancing security level and time for successful signing, where $w$ is larger than the error correctability $t$. To determine what is the appropriate value of $w$ in the proposed signature scheme, we perform simulations for random syndromes. For $N$ random syndromes $s$, we find the minimum Hamming weight error vector $e$ satisfying $H'e^T = s$ by carrying out complete decoding. The required number $N$ of counters $i$, the corresponding error weight parameter $w$, and probability of successful signing in the signing stage are listed in Table I.

\begin{figure}
        \begin{center}
        \includegraphics[width=0.7\columnwidth,keepaspectratio]
        {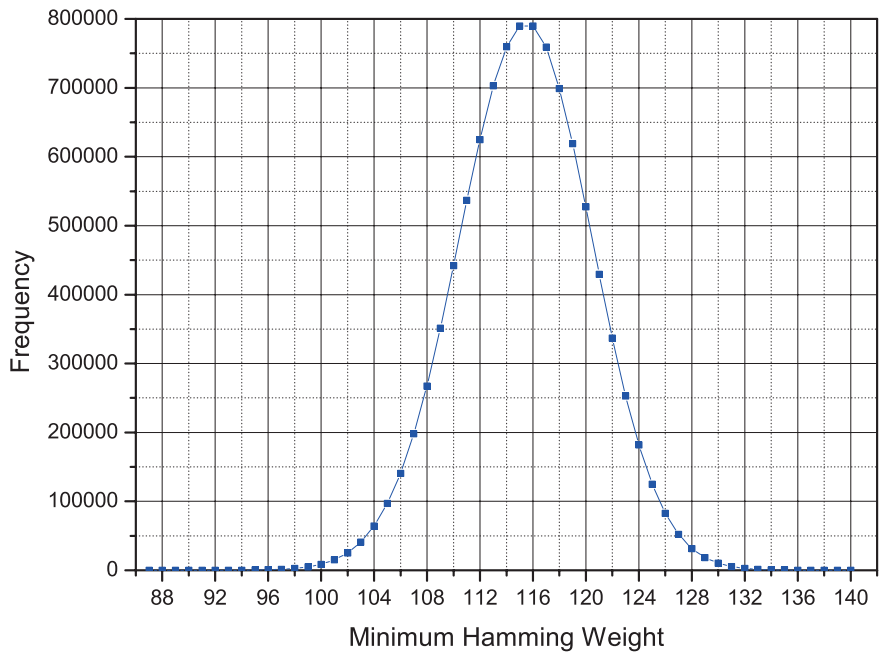}%
        \caption{Distribution of Hamming weights of coset leaders among $10^7$ in RM$(10,5)$.}
        \label{fig_4}
        \end{center}
\end{figure}

Assume that $N$ is the maximum number of signing trials for the successful signing in the signing stage. The signing is successful if the complete decoded error weight is less than or equal to $w$ for the hashed message with counter $i$, $h(h(M)|i)$. Let $X_i$ be the Hamming weight of error vector by the complete decoding for counter $i$. Then the probability of successful signing is given as
\begin{align}
{\rm prob}\left\{\min_{i\le N}(X_i) \le w\right\} &= 1 - {\rm prob}\left\{X_1 > w, X_2 > w, {\cdots}, X_N > w \right\} \nonumber \\
&= 1 - \left({\rm prob}\{X_1 > w\}\right)^N \label{13}
\end{align}
where each $X_i$ is assumed to be i.i.d. The probability ${\rm prob}\{X_1 > w\}$ can be numerically obtained by the distribution of Hamming weights of coset leaders in the complete decoding. Thus, $N$ and $w$ can be selected for successful signing for the given RM code using (6). For RM$(10,5)$, the distribution of Hamming weights of coset leaders is numerically obtained from Fig.~\ref{fig_4}, where the minimum Hamming weight of error vectors among $10^7$ random syndromes is 87. In Table I, the probability of successful signing for RM(10,5) is listed for parameters $N$ and $w$ using (\ref{13}).

\begin{table}[t]
\caption{The probability of successful signing for parameters $N$ and $w$ in RM$(10,5)$.}
\begin{center}
\begin{tabular}{|c|c|c|c|}
\hline
$w\setminus N$ &10,000&20,000&40,000 \\ \hline
90 & 0.01 &0.02 & 0.04 \\ \hline
93 & 0.1 &0.18 & 0.33 \\ \hline
96 & 0.41 &0.83 & 0.97 \\ \hline
97 & 0.83 &0.97 & 1 \\ \hline
98 & 0.97 &1 & 1 \\ \hline
99 & 1 & 1& 1 \\ \hline
\end{tabular}
\end{center}
\end{table}

\begin{table}
\caption{An error weight parameter $w$ and their signing time for 10,000 random syndromes.}
\begin{center}
\begin{tabular}{|c|c|c|c|}
\hline
&$(n,k,d_{\rm min})$&\makecell{Error weight parameter $w$ \\ for successful signing}& \makecell{Successful \\signing time}\\ \hline
RM(10, 4) &(1024, 386, 64)& 192 & 2.26 sec  \\ \hline
RM(10, 5) &(1024, 638, 32)& 97 & 2.21 sec \\ \hline
RM(11, 5) &(2048, 1024, 64)& 306 & 4.45 sec \\ \hline
RM(12, 5) &(4096, 1586, 128)& 855 & 9.50 sec \\ \hline
RM(12, 6) &(4096, 2510, 64)& 458 & 8.94 sec \\ \hline
\end{tabular}
\end{center}
\end{table}

From Table I, using the parameter $w = 99$, the probability of successful signing is almost 1 for $N = 10000$ in RM(10, 5). If we simulate for more syndromes, smaller weight of $e$ can be obtained at the cost of longer signing time. That is, if we choose the smaller error weight parameter $w$ and the larger $N$ in the proposed signature scheme, the level of security becomes higher, but the time for successful signing is increased. In Table II, the error weight parameter $w$ that can be successfully signed with $N=10,000$ trials and its signing time in a straightforward implementation on Intel i7 Processor of 3.0 GHz for each RM code are described. 

The proposed signature scheme is summarized as in Algorithm \ref{summary}.

\begin{algorithm}
\caption{The proposed signature scheme}
\label{summary}
Preprocessing:
	\begin{list}{ }{\setlength{\leftmargin=5mm}}
		\item For a given modified $(n,k)$ RM code and the security level larger than 128 bits, derive $(N,w)$ for successful signing as in Table I.
	\end{list}
Key Generation:
	\begin{list}{ }{\setlength{\leftmargin=5mm}}
		\item Generate random matrices $S$, $Q$, and $R$.
		\item Generate $H_m$ as in Algorithm 2.
		\item Compute $H'=SH_mQ$.
	\end{list}
Signing:
	\begin{list}{ }{\setlength{\leftmargin=5mm}}
		\item For $i = 1$ to $N$
		\begin{list}{ }{\setlength{\leftmargin=5mm}}
			\item Find syndrome $s=h(h(M)|i)$ and compute $s'=S^{-1}s$.
			\item Find $e'$ such that $He'^T = s'$ by complete decoding.
			\item Find $e'_p = s'_p + Re_{n-p}^T$ and thus $e' = [e'_{n-p}|e'_p]$.
			\item If ${\rm wt}(e') \le w$, then goto *.
		\end{list}	
		\item end
		\item * Compute $e^T = Q^{-1}e'^T$ and thus signature is $\sigma = (M,e,i)$.
	\end{list}
Verification:
	\begin{list}{ }{\setlength{\leftmargin=5mm}}
		\item Check ${\rm wt}(e') \le w$ and $H'e^T = h(h(M)|i)$.
		\item If True, then return ACCEPT, else return REJECT.
	\end{list}	
\end{algorithm} 

\vspace{10pt}
\section{{Security Analysis of the Proposed Code-Based Signature Scheme}}
\subsection{EUF-CMA}
In this subsection, we prove that the proposed signature scheme is secure under EUF-CMA. The proposed signature scheme can be simplified as follows. We can consider finding $e$ satisfying $H'e^T = h(h(M)|i)$ as a signing process, where the parity check matrix of a linear code is $H_{m}$, and $\gamma$ is the decoding algorithm in the signature scheme in the previous section. Then, the proposed signature scheme can be considered to be the same as the original CFS scheme for the EUF-CMA security check.

\vspace{0.1in}
{\bf 1) Key Generation:}

Let $H_{m}$ be a parity check matrix of a modified RM code with the decoding algorithm $\gamma$.
Then, the private keys are $S$, $Q$, and $H_{m}$, and the public keys are $H' = SH_{m}Q$ and $w$.

\vspace{0.1in}

{\bf 2) Signing:}

\begin{list}{ }{\setlength{\leftmargin=5mm \labelwidth=1mm}}
\item To sign a message $M$,
\item For $i = 1$ to $N$
\begin{list}{ }{\setlength{\leftmargin=5mm}}
	\item $e' = \gamma (S^{-1}h(h(M)|i))$
	\item if ${\rm wt}(e') \le w$, go to *.
\end{list}
\item end
\item * Output $(M,  e = e'(Q^{-1})^T, i)$
\end{list}

\vspace{0.1in}

{\bf 3) Verification}

Check ${\rm wt}(e) \le w$ and $H'e^T = h(h(M)|i)$. If TRUE, then return ACCEPT; otherwise, return REJECT.

To prove the EUF-CMA security, we need the following assumption and proposition. The differences between the proposed signature scheme and the CFS signature scheme are; i) the use of a different code, namely, a modified RM code rather than a Goppa code, and ii) the use of complete decoding instead of syndrome decoding.

\begin{assumption}[RM code distinguishability problem]
There is no probabilistic polynomial time (PPT) distinguisher ${\mathcal D}$ that can distinguish $H'=SH_{m} P$ from a randomly generated parity check matrix $H_R$.
\end{assumption}

To the best of our knowledge, there are no known algorithms for distinguishing a modified parity check matrix $H'$ of an RM code from $H_R$ up to now, and thus we set the following assumption.

\begin{proposition}[\cite{NPHard} Hardness of complete decoding]
The complete decoding problem for an $(n,k,t)$ linear code is an NP-complete problem if the Hamming weight of $e$ is less than $n/3$.
\end{proposition}

With this assumption and proposition, the following theorem holds.
\begin{theorem}
The proposed modified RM code-based signature scheme is EUF-CMA secure.
\end{theorem}
\begin{proof}
The proof of Theorem 7 is provided in Appendix.
\end{proof}

\subsection{Security by Error Weight Parameter}
The attacker tries to forge the signature with public key $H'$ and hashed message $h(h(M)|i)$. Assume that the public key $H'$ is systematic, where $H' = [H_0|I]$, $I$ is an $(n-k)\times (n-k)$ identity matrix, and $H_0$ is an $(n-k)\times k$ matrix. Then, attacker computes $z$ satisfying the following equation
\begin{equation}
H'z^T=[H_0|I][z_1|z_2]^T = s = h(h(M)|i)
\end{equation}
where $z_1$ and $z_2$ are vectors with size $k$ and $n-k$, respectively. The attacker can let $z_1$ be an all-zero vector and $z_2 = s$. If the Hamming weight of $z_2$ is less than or equal to $w$, then the forgery is successful. The probability of successful forgery is given as

\begin{equation}
\frac{\sum_{i=0}^{w}{n-k \choose i}}{2^{n-k}},
\end{equation}
which is presented in Table III for each RM code and error weight parameters $w$ of error vectors given in Table I.
However, for RM(11, 5) with $w=300$, the probability of successful forgery is less than $2^{-128}$ but it requires $N = 200,000$ for successful signing. Then, the proposed signature scheme using the RM(11, 5) code with $w=300$ is considered to be 128 bit-secure.

\begin{table}
\caption{The security of the proposed signature scheme for $N=10,000$.}
\begin{center}
\begin{tabular}{|c|c|c|}
\hline
&$(n,k,d_{\rm min}, w)$& $\frac{\sum_{i=0}^{w}{n-k \choose i}}{2^{n-k}}$ \\ \hline
RM(10, 4) &(1024, 386, 64, 192)&  $\le 2^{-74}$  \\ \hline
RM(10, 5) &(1024, 638, 32, 98)& $\le2^{-70}$ \\ \hline
RM(11, 5) &(2048, 1024, 64, 306)& $\le2^{-122}$  \\ \hline
RM(12, 5) &(4096, 1586, 128, 855)& $\le2^{-186}$  \\ \hline
RM(12, 6) &(4096, 2510, 64, 458)& $\le2^{-209}$  \\ \hline
\end{tabular}
\end{center}
\end{table}

\vspace{10pt}
\section{Conclusions}
In this paper, we proposed a new signature scheme based on the punctured RM code with random insertion. The proposed signature scheme improves the Goppa code-based CFS signature scheme by increasing the probability of successful signing using the complete decoding method. In addition, the proposed signature scheme can avoid some known attacks for the RM code-based cryptosystem using the puncturing method with random insertion. The optimal parameters of the signing time and security were derived. It was also proved that the proposed signature scheme achieves EUF-CMA security.

\vspace{10pt}
\section*{Acknowledgement}
This work was supported by the Institute for Information \& communications Communications Technology Promotion (IITP) grant funded by the Korea government (MSIP) (R-20160229-002941, Research on Lightweight Post-Quantum Crypto-systems for IoT and Cloud Computing).
\vspace{10pt}
\vspace{10pt}
\section*{Appendix}
{\bf Proof of Theorem 7:}
The proof of this theorem is almost the same as the proof of the EUF-CMA security of the CFS signature scheme \cite{EUFCMA}. {It was proved in \cite{EUFCMA} that a variant of a CFS scheme is EUF-CMA secure if certain assumptions are true. However, it was shown that the assumption that distinguishing Goppa codes from random codes is difficult is not true for the case of some parameters (small $t$) used in the CFS signature scheme. Thus, we can follow the logic of the proof in \cite{EUFCMA} because the adopted assumption for the proposed modified RM code-based signature scheme is still valid.}
We define the sequence of games $G_0, G_1, {\cdots}$, $G_5$ in the same way as in \cite{EUFCMA}. Let $G_0$ be the {original} security game, that is, the EUF-CMA game, and $G_5$ be solving the syndrome decoding problem.

The main differences between the proposed signature scheme and the CFS signature scheme are mostly in Games $G_3$ and $G_5$. {In the proof of the CFS signature scheme,} Game $G_3$ discusses the Goppa code distinguishing problem, but for a small $t$, it turns out to be distinguishable from a random code. In the case of the proposed signature scheme, we adopt the modified RM code distinguishing problem in Assumption 5 because there has been no way to prove the distinguishability up to now. {Although Game $G_5$ is related to the syndrome decoding problem in the proof of the original CFS signature scheme, we will} replace it using {the} complete decoding problem, {which is known as an NP-complete problem.} A full description of this proof is given as follows.

The challenger ${\mathcal C}$ plays a sequence of games {$G_0, G_1$}, ${\cdots}$, {$G_5$}. Here, {$G_0$ corresponds to the standard} EUF-CMA game as mentioned above. In {$G_0$}, the adversary ${\mathcal A}$ tries to forge a signature. If the adversary ${\mathcal A}$ successfully forges the signature, then ${\mathcal A}$ wins the game {$G_0$}. Successive games are given through slight modifications of the preceding games. Let ${\rm Pr}[G_i]$ be the winning probability of each game $G_i$. We then have to prove that the probability of the winning condition of these games is proved to be arbitrarily small through all of the intermediate games.

{$G_0$}: The challenger ${\mathcal C}$ obtains the private and public keys using a key generation algorithm. The adversary ${\mathcal A}$ obtains the public key $H'$, and can access a hash oracle ${\mathcal H}$ and signing oracle $\Sigma$. Let {$q_h$ and $q_s$} be the maximum {numbers} of queries made by the adversary ${\mathcal A}$ to {the hash oracle and the signing oracle, respectively.} The procedure of {$G_0$} is given in Algorithm \ref{game0}. Then, the winning probability of {$G_0$} is given as

\begin{equation}
\label{g0}
{\rm Pr}[G_0]={\rm succ}^{\rm EUF-CMA}({\mathcal A}).
\end{equation}

\begin{algorithm}
\caption{$G_0$ (EUF-CMA)}
\label{game0}
\begin{enumerate}
	\item $(H', S, H, Q) \leftarrow {\rm keygen}({\mathcal C})$
	\item Set the oracles ${\mathcal H}$ and $\Sigma$
	\item $(M^*, \sigma^*, i^*) \leftarrow {\mathcal A}^{{\mathcal H},\Sigma}(H')$
	\item If ${\mathcal H}(M^*, i^*) = H'\sigma^{*T}, {\rm wt}(\sigma^*)\le w,$ and $\Sigma$ did not provide $\sigma^*$, then
	\begin{list}{ }{\setlength{\leftmargin=5mm}}
		\item ${\mathcal A}$ wins the game
	\end{list}
	else
	\begin{list}{ }{\setlength{\leftmargin=5mm}}
		\item ${\mathcal A}$ loses the game
	\end{list}
	end
\end{enumerate}
\end{algorithm}

{$G_1$}: In this game, the challenger modifies {the} hash oracle ${\mathcal H}$ by ${\mathcal H'}$. In ${\mathcal H'}$, the challenger uses {a list} $\Lambda$ that consists of counter {values} of $i=\Lambda(M)$ for message $M$ such that ${\mathcal H}(M, i)$ is a decodable syndrome {and another list $\Lambda_{\mathcal H}$ to store a valid syndrome-error pair that was already produced in the previous queries. If there is no element corresponding to the input, the output is $\perp$.} The modified hash oracle ${\mathcal H'}$ produces syndromes {according to} Algorithm \ref{game1}, and finally produces $q_h+q_s+1$ syndromes. In addition, it is known that the relation of ${\rm Pr}[G_0]$ and ${\rm Pr}[G_1]$ is given as
\begin{equation}
\label{g1}
|{\rm Pr}[G_1]-{\rm Pr}[G_0]| \le \epsilon_0
\end{equation}
where $\epsilon_0 = 1 - \left(1-\frac{1}{2^{n-k}}\right)^{q_h+q_s+1}$ \cite{EUFCMA}.

\begin{algorithm}
\caption{Game $G_1$ (${\mathcal H'}$: simulation of ${\mathcal H}$)}
\label{game1}
Input: a pair $(M,i)$\\
Output: a syndrome $s$
\begin{enumerate}
	\item If $\Lambda(M) = \perp$, then\\
		$\Lambda(M) \xleftarrow{R} \{1,{\cdots}, 2^{n-k}\}$
	\item $(s,e) \leftarrow \Lambda_{\mathcal H}(M,i)$
	\item If $i\neq \Lambda(M)$, then
	\begin{list}{ }{\setlength{\leftmargin=5mm}}
		\item  If $s=\perp$, then
		\begin{list}{ }{\setlength{\leftmargin=5mm}}
			\item $s \xleftarrow{R} F_2^{n-k}$
			\item $\Lambda_{\mathcal H}(M,i) \leftarrow (s,\perp)$
		\end{list}
		\item end
		\item return ${\mathcal H}(M,i)=s$
	\end{list}
	else
	\begin{list}{ }{\setlength{\leftmargin=5mm}}
	\item If $s=\perp$, then
		\begin{list}{ }{\setlength{\leftmargin=5mm}}
			\item $e \xleftarrow{R} \{y\in F_2^n|{\rm wt}(y) \le w \}$
			\item $s \leftarrow He^T$
			\item $\Lambda_{\mathcal H}(M,i) \leftarrow (s,e)$
		\end{list}
	\item end
	\item return ${\mathcal H'}(M,i)=s$
	\end{list}
	end
\end{enumerate}
\end{algorithm}

{$G_2$}: In {$G_2$}, the {challenger} replaces the signing oracle $\Sigma$ with ${\Sigma'}$. {The modified signing} oracle queries ${\mathcal H'}$ on $(M,\Lambda(M))$ {according to} Algorithm \ref{game2}. In addition, the winning probability relation of $G_1$ and $G_2$ is derived as
\begin{equation}
\label{g2}
|{\rm Pr}[G_2]-{\rm Pr}[G_1]| \le \epsilon_1
\end{equation}
where $\epsilon_1 = 1 - \left(1-\frac{q_s}{2^{n-k}}\right)^{q_h}$.

\begin{algorithm}
\caption{Game $G_2$ ($\Sigma'$: simulation of $\Sigma$)}
\label{game2}
Input: a message $M$\\
Output: a signature $(i,\sigma)$
\begin{enumerate}
	\item If $\Lambda(M) = \perp,$ then
	\begin{list}{ }{\setlength{\leftmargin=5mm}}
		\item $\Lambda(M) \xleftarrow{R} \{1,{\cdots}, 2^{n-k}\}$
	\end{list}
	end
\item ${\mathcal H'}(M,\Lambda(M))$
\item $(s,x) \leftarrow \Lambda_{{\mathcal H'}}(M,\Lambda(M))$
\item $\Lambda(M) \leftarrow \perp$
\item Return $\Sigma(M) = (i, x)$
\end{enumerate}
\end{algorithm}

{$G_3$}: In {$G_3$}, the challenger replaces the key generation algorithm with the selection of a random binary parity check matrix. The selected parity check matrix is taken as the public key. {Because neither} the hash oracle nor the signature oracle {uses the hash function and the private keys}, the difference in the winning probabilities of $G_2$ and $G_3$ is the same as the distinguishing probability between the modified RM code and a random binary code, that is,
\begin{equation}
\label{g3}
|{\rm Pr}[G_3]-{\rm Pr}[G_2]| \le \epsilon_{\rm distinguish}.
\end{equation}
By Assumption 5, the value of $\epsilon_{\rm distinguish}$ is negligible. {The} description of {$G_3$} is given as Algorithm \ref{game3}.

\begin{algorithm}
\caption{Game $G_3$}
\label{game3}
Input: a parity check matrix $H$\\
Output: a bit $b$
\begin{enumerate}
	\item Given $w$, set the oracles ${\mathcal H'}$ and $\Sigma'$
	\item $(M^*, \sigma^*, i^*) \leftarrow {\mathcal A}^{{\mathcal H'},\Sigma'}(H) $
	\item If ${\mathcal H'}(M^*, i^*) = H\sigma^{*T}, {\rm wt}(\sigma^*)\le w$, and $\Sigma'$ did not provide $\sigma^*$, then
	\begin{list}{ }{\setlength{\leftmargin=5mm}}
		\item $b = 1$
	\end{list}
	else
	\begin{list}{ }{\setlength{\leftmargin=5mm}}
		\item $b = 0$
	\end{list}
	end
\end{enumerate}
\end{algorithm}

{$G_4$}: {$G_4$} is conditioned by an adversary making a forgery on a particular hash query. The challenger first obtains a random $c \xleftarrow{R} \{1,{\cdots},q_h+q_s+1\}$. Adversary ${\mathcal A}$ wins the game if the $c$th query to ${\mathcal H'}$ is made on $(M^*, i^*)$. Because $c$ is randomly chosen from $q_h+q_s+1$ possibilities, the winning probability of $G_4$ is given as
\begin{equation}
\label{g4}
{\rm Pr}[G_4] = \frac{{\rm Pr}[G_3]}{q_h+q_s+1}.
\end{equation}

{$G_5$}: In this game, the challenger {modifies} the hash oracle to output a random syndrome $s^*$ to the $c$th query. The {winning} probability of $G_5$ is the same as the winning probability of {$G_4$}. {The detailed procedure for $G_5$} is given in Algorithm \ref{game5}. {Note that} this game is the same as solving the complete decoding problem. Then,
\begin{equation}
\label{g5}
{\rm Pr}[G_5] = {\rm Pr}[G_4] \le \epsilon_{\rm complete}.
\end{equation}
From Proposition 6, the value of $\epsilon_{\rm complete}$ is negligible.

\begin{algorithm}
\caption{Game $G_5$}
\label{game5}
Input: an adversary ${\mathcal A}$
\begin{enumerate}
	\item $c \xleftarrow{R} \{1,{\cdots}, q_{\mathcal H}+q_{\Sigma}+1\}$
	\item $H^* \xleftarrow{R} (n,k)$ binary code for given $w$
	\item $s^* \xleftarrow{R} F_2^{n-k}$
	\item Set the oracles ${\mathcal H'}$ and $\Sigma'$
	\item $(M^*, \sigma^*, i^*) \leftarrow {\mathcal A}^{{\mathcal H'},\Sigma'}(H^*)$
	\item If $
	\begin{cases}
		{\mathcal H'}(M^*, i^*) = H\sigma^{*T}\\
		{\rm wt}(\sigma^*)\le w \\
	\end{cases}$ and $
	\begin{cases}
		\Sigma' \;{\rm did \;not \;provide\; \sigma^*}\\
		c-{\rm th\; query\; to\;} {\mathcal H'} \;{\rm was}\; (M^*, i^*),
	\end{cases} $ \\
	then
	\begin{list}{ }{\setlength{\leftmargin=5mm}}
		\item ${\mathcal A}$ wins the game
	\end{list}
	else
	\begin{list}{ }{\setlength{\leftmargin=5mm}}
		\item ${\mathcal A}$ loses the game
	\end{list}
	end
\end{enumerate}
\end{algorithm}

Combining all of the above equations, (\ref{g0})--(\ref{g5}), we have
\begin{equation}
{\rm succ}^{\rm EUF-CMA}({\mathcal A}) \le (q_h+q_s+1)\epsilon_{\rm complete} + \epsilon_{\rm distinguish} + \epsilon_0 + \epsilon_1.
\end{equation}
Hence, the probability of a successful forgery is negligible if the punctured RM codes with random insertions are indistinguishable from random linear codes and the complete decoding problem is intractable. Thus, the proposed signature scheme is EUF-CMA secure.

\hfill $\Box$

%








\end{document}